\documentstyle[prl,aps,epsf,rotate,multicol]{revtex}

\begin{document}
\draft


\begin{multicols}{2}


{\bf Comment on: `Universal Behavior of Load Distribution in
Scale-free Networks'}

\begin{figure}
\narrowtext
\centerline{
\epsfysize=0.4\columnwidth{\epsfbox{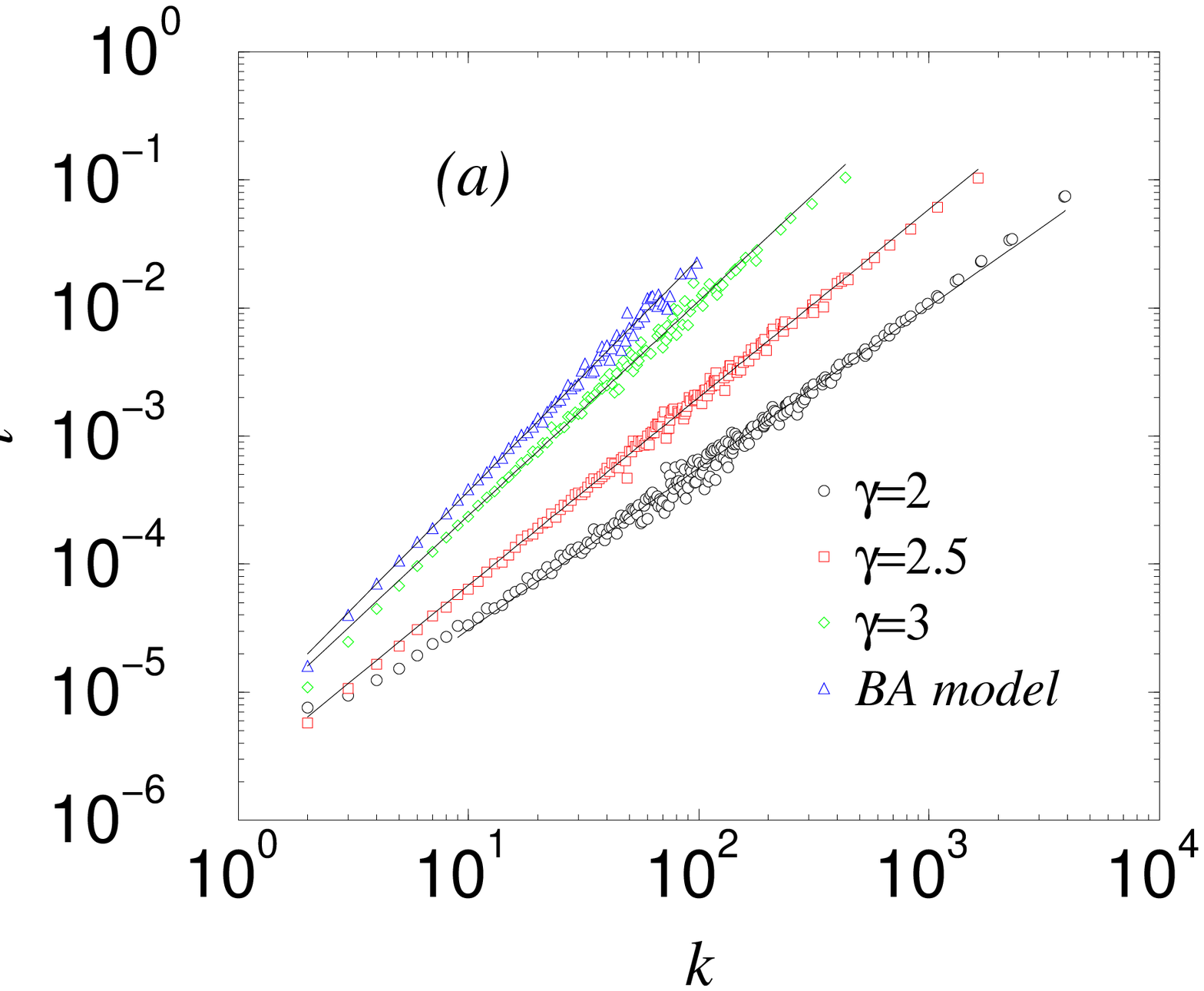}}
\hspace{0.1cm}
\epsfysize=0.4\columnwidth{\epsfbox{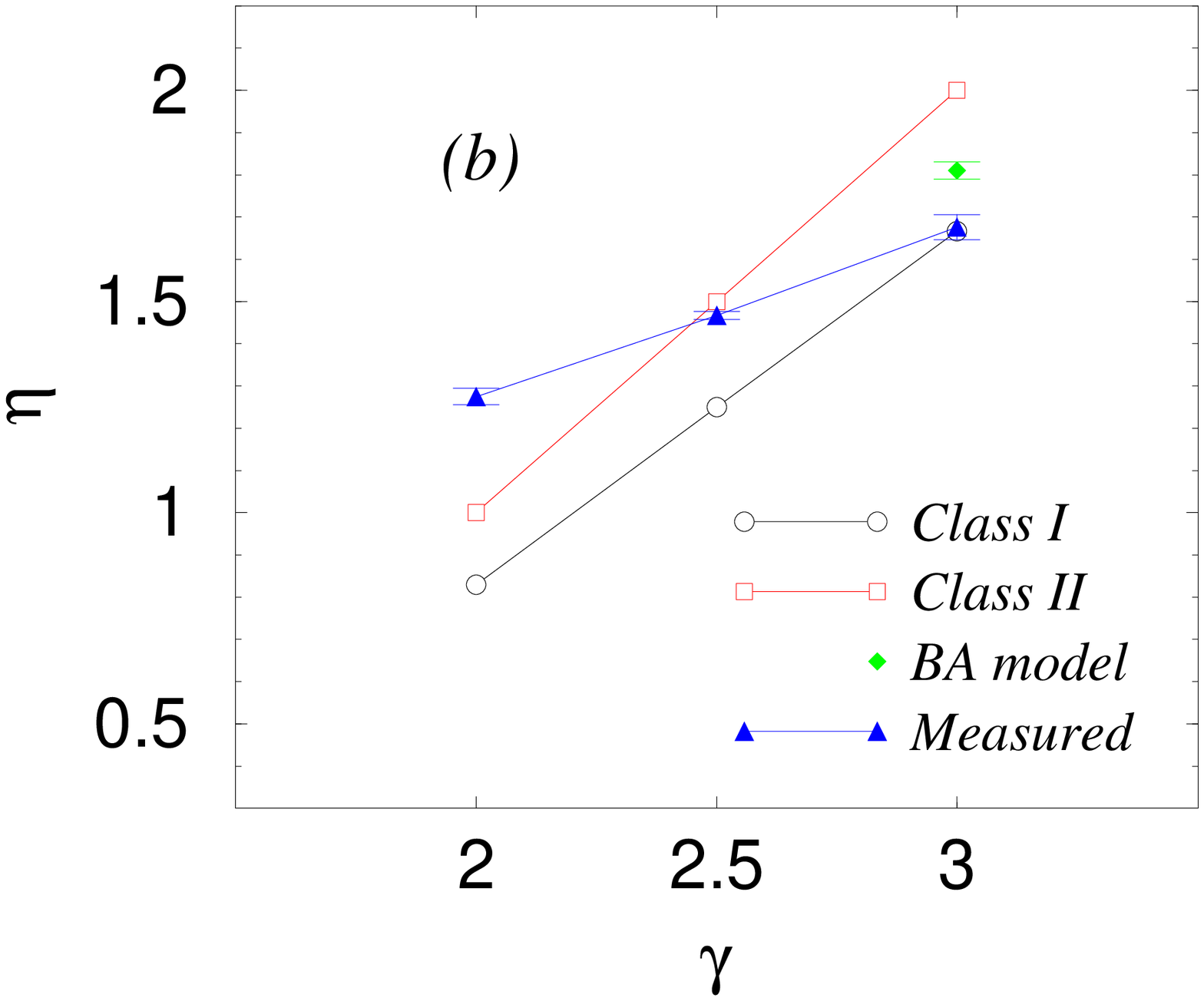}}
}
\vspace*{0.2cm}
\caption{ (a) Log-Log plot of the normalized average load versus
connectivity for the same models as in {\protect\cite{Goh:2001}} with
$m=2$. The power law fits (straight lines) give $\eta=1.27\pm0.01$
($N=3.10^4$), $1.467\pm0.006$ ($N=5.10^4$), and $1.68\pm0.02$
($N=5.10^4$) for $\gamma=2$, $2.5$, and $3$ respectively. For the BA
model, $\eta=1.81\pm0.02$ ($N=5.10^4$). (b) {\protect{$\eta$}} versus
{\protect{$\gamma$}}. If the universality proposed in
{\protect\cite{Goh:2001}} would be correct, the measured values for
$\gamma\in[2,3)$ should lie on the ``universal'' straight line
corresponding to $\delta=2.2$ (class I).}
\label{bc.and.univ}
\end{figure}

In a previous Letter\cite{Goh:2001}, Goh {\it et al} have presented a
numerical study of the load---or betweenness centrality---distribution
in a scale-free network whose degree distribution follows a power law
$p(k)\sim k^{-\gamma}$ where $\gamma\in [2,\infty[$ is a tunable
parameter. They showed that the load $\ell$ is distributed according
to a power-law $P(\ell)\sim\ell^{-\delta}$ with exponent $\delta$. On
the basis of their numerical results, they conjectured that the value
of $\delta\simeq 2.2$ is independent of $\gamma$ for the interval
$[2,3)$. Based on this apparent universality, a classification of
scale-free networks according to the value of $\delta\simeq 2.2$
(class I) or $\delta=2$ (class II) was proposed\cite{GohPNAS:2002}. In
this comment we argue that the value of $\delta$ is not universal and
varies significantly as $\gamma$ changes in the interval $[2,3)$. The
power law fits of the cumulative function $\text
{Prob}(\text{load}\ge\ell)$ for the model proposed in\cite{Goh:2001}
gives the values $\delta=1.84\pm0.04$, $2.05\pm0.05$ and $2.25\pm0.05$
for $\gamma=2$, $2.5$, and $3$ respectively, while for the BA
model\cite{Barabasi} $\delta=2.3\pm0.1$. The variations of $\delta$
are significant enough to claim that it is not universal but in order
to double-check our results we use an indirect way of computing
$\delta$. We study the relation between the load and the
connectivity\cite{Goh:2001,Vasquez} which is of the form $\ell\sim
k^\eta$ where the exponent $\eta$ depends on the network. As can be
seen on Fig.~(\ref{bc.and.univ}a), the power law holds remarkably for
a large range of $k$ allowing for an accurate measure of $\eta$. We
also checked that the value of $\eta$ does not change significantly
for different values of the system size (For $\gamma=2.5$, we obtain a
relative variation due to size going from $N=10^4$ to $5.10^4$ less
than $1\%$). The exponents $\eta$ and $\delta$ are not independent and
it is easy to show that\cite{Vasquez} $\eta=(\gamma-1)/(\delta-1)$. If
the value of $\delta\simeq 2.2$ is universal then $\eta$ is a linear
function of $\gamma$ with slope $\simeq 1/1.2\simeq 0.83$. In
Fig.~(\ref{bc.and.univ}b) we plot the measured $\eta$ versus $\gamma$
for the different types of networks studied and the corresponding
value predicted by universality. This Fig.~(\ref{bc.and.univ}b) shows
that if for $\gamma\simeq 3$ the value $\delta=2.2$ seems to be
acceptable, the claim of universality for $\gamma\in [2,3)$ proposed
in\cite{Goh:2001} does not hold (our results do not fit in the other
class $\delta=2.0$ either). In addition, we tested the universality
for different values of $m$ and we also obtain variations ruling it
out: For $\gamma=2.5$ and for $N=2.\/10^4$, we obtain
$\eta=1.477\pm0.006, 1.56\pm0.006$, and $1.64\pm0.01$ for $m=2,4,6$
respectively.

The important exponent thus appears to be $\eta$ and it is interesting
to note that $\eta$ is significantly smaller than the maximum value
$\eta=2$. This maximum value is reached when nodes with large
centrality (ie. with large $k$) link together disconnected parts of
roughly the same size.  The load of these nodes is then of the order
of $\ell\sim k(k-1)/2\sim k^2$. The fact that $\eta<2$ indicates that
the different parts are also connected by shortest paths which do not
pass through the central node. More generally, it would be interesting
to understand how $\eta$ depends on the different parameters of the
network such as $\gamma$ and the degree correlation.

\smallskip
\indent{Marc Barth\'elemy}\\
\indent{CEA-Service de Physique de la Mati\`ere Condens\'ee}\\
\indent{BP12 Bruy\`eres-Le-Ch\^atel, France}


\end{multicols}



\begin{references}

\bibitem{Goh:2001} 
K.-I.~Goh, B.~Kahng, and D.~Kim, Phys. Rev. Lett. {\bf 87}, 278701
(2001).

\bibitem{GohPNAS:2002} 
K.-I. Goh, H. Jeong, B. Kahng, and D. Kim,
Proc. Natl. Acad. Sci. (USA) {\bf 99}, 12583 (2002).


\bibitem{Barabasi}
A.-L.~Barabasi and R.~Albert, Science {\bf 286}, 509 (1999).

\bibitem{Vasquez} A.~Vazquez, R.~Pastor-Satorras, and A.~Vespignani,
Phys. Rev. E {\bf 65}, 066130 (2002).

\end{references}
\end{document}